\documentclass[aps,preprint,showpacs,floatfix]{revtex4-1}

\usepackage{graphicx}
\usepackage{amssymb}
\usepackage{amsmath}
\usepackage{subfig}
\usepackage{float}

\def\bea{\begin{eqnarray}}
\def\eea{\end{eqnarray}}
\def\beq{\begin{equation}}
\def\eeq{\end{equation}}

\def\bm{\begin{math}}
\def\me{\end{math}}

\begin{document}
\title{Velocity Distribution of a Uniformly Heated Hard Sphere Granular Gas}
\author{Rameez Farooq Shah} \email{rmzshah@gmail.com}
\affiliation{Department of Physics, Jamia Millia Islamia (A Central University), New Delhi 110025, India}
\author{Shikha Kumari} \email{shikha1011@gmail.com}
\affiliation{Department of Physics, School of Basic and Applied Sciences, IILM university, Greater Noida, Uttar Pradesh 201306}
\author{Syed Rashid Ahmad} \email{srahmad@jmi.ac.in}
\affiliation{Department of Physics, Jamia Millia Islamia (A Central University), New Delhi 110025, India}

\begin{abstract}
This paper presents a molecular dynamics simulation of an inelastic gas, where collisions between
molecules are characterized by a coefficient of restitution less than unity. The simulation employs
an event-driven algorithm to efficiently propagate the system in time, tracking molecular positions
and velocities. A thermostat mechanism is incorporated to maintain the system's temperature
by applying Gaussian white noise to the molecular velocities. The system's kinetic energy evolves 
towards a non-equilibrium steady state, with the initial dynamics governed by the interplay between 
energy input from the thermostat and energy dissipation through inelastic collisions. This steady 
state emerges when the energy gain from the thermostat balances the energy loss due to inelastic 
collisions. We calculate the coefficients of the Sonine polynomial expansion of the velocity 
distribution function to show that the velocity distribution exhibits a departure from the 
Maxwell-Boltzmann distribution in the steady state.
\end{abstract}
\maketitle
\section{\label{introd} INTRODUCTION}
The study of granular materials, which are systems consisting of macroscopic particles interacting through dissipative collisions, has attracted significant attention in the past few decades due to their widespread applications in various industries and natural phenomena. These materials exhibit novel properties similar to those of fluids and solids \cite{rmp_behringer, rmp_kadanoff}. As solids, they form heaps and withstand deformation. A pile of sand at rest is an example. However, dry sand or powders also flow through the neck of an hourglass like a liquid. Dry sand can also be agitated in some external drive geometry 
so as to behave like a gas. The constituent particles of a granular material are polydisperse in size and shape; usually larger than $1 \mu m$ in size and arbitrary shapes. The macroscopic size essentially means that they are not subject to thermal fluctuations. In theoretical and numerical studies, they are usually modelled as spheres, needles or cylinders \cite{rmp_tsimring, duran, ristow, nb_ktgg}. One of the most distinctive properties of granular materials is the dissipative interaction between constituent particles. The interactions result in a loss of kinetic energy or \textit{cooling}, accompanied  by a local parallelization of particle velocities. Because of the dissipative nature of particle–particle interactions, many interesting phenomena like size separation, clustering, pattern formation, inelastic collapse, anomalous velocity statistics etc., have been reported \cite{haff83, swinney9596, gz93, mcny9296, jjbrey9698, tpcvn9798, sl9899, ap0607, adsp1213}.

The granular gas (dilute granular systems) is a paradigm to understanding the properties of a gas whose molecules dissipative energy as a result of molecular interaction. The starting point in the study of a granular gas is the evolution of a homogeneously distributed inelastic particles. In the absence of energy input from an external source, the system loses its kinetic energy due to inelastic collisions between particles. Initially, the density appears homogeneous and the system loses energy in a \textit{homogeneous cooling state} (HCS). However, due to fluctuations in the density and velocity fields, the HCS is unstable and the system evolves into an inhomogeneous cooling state (ICS) \cite{haff83, ap0607, dp03}. In the ICS, regions of particle rich clusters emerge and grow  continuously and particles in a cluster start moving in approximately parallel directions. In a typical experimental setting, the loss of energy is often compensated by energy input through various drive geometries like horizontal or vertical vibration or rotation. The system, in these situations, settles into a nonequilibrium steady state \cite{swinney9596, ristow}. 

The study of granular materials, which are systems consisting of macroscopic particles 
interacting through dissipative collisions, has attracted significant attention in the past few  decades due to their widespread applications in various industries and natural phenomena. 
These materials exhibit novel properties similar to those of fluids and solids [1, 2]. 
The choice of thermostat mechanism plays a crucial role in determining the statistical 
properties of driven granular systems. Different driving mechanisms can lead to distinct 
steady-state behaviors and velocity statistics. In this work, we focus specifically on the 
white-noise thermostat, which provides uniform heating throughout the system, while acknowledging that other thermostating methods may yield different results.

An important problem in the context of a granular gas is the study of velocity distribution of a granular (or inelastic) gas. It is known for more than a century that the steady state velocity distribution of a gas with elastic molecular interaction is the Maxwell-Boltzmann (MB) distribution. This is not the case with inelastic or granular gases. Studies on granular gases have shown a departure from the MB distribution \cite{ap0607, adsp1213, pdsp2018}. In this paper, we study the velocity distribution of a hard sphere granular gas which has been heated uniformly using a Gaussian white noise thermostat. The departure from MB distribution is characterized by calculating the coefficients of the Sonine polynomials expansion. In analytical studies and computer simulations, heated granular gases have been studied extensively \cite{vne98}. In order to inject energy to the system, a thermostat mechanism is usually employed. In our studies, we apply the algorithm suggested Williams et al., where a white-noise thermostat (WNT) is used to heat the particles uniformly \cite{willmac96, william96}. 

The paper is organized as follows. In Sec.\ref{themodel}, we discuss the details of our model of a uniformly heated granular gas. The characterization of the velocity distribution function in terms of the coefficients of the Sonine polynomial expansion is discussed in Sec.\ref{vdf}. In Sec.\ref{details_results}, we present detailed results from our molecular dynamics simulations, focusing on the velocity distribution of a heated granular gas. Finally, Sec.\ref{sec:summry} summarizes our findings and discusses their implications for understanding the velocity statistics of driven granular systems.

\section{\label{themodel} MODEL}
Our starting point is a homogeneous granular gas, consisting of identical spherical molecules. Without loss of generality we may assume mass and diameters of the molecules to be unity. For hard sphere molecules, the post-collision velocities of the particles labeled as $i$ and $j$ are given as a function of pre-collision velocities by the rule:
\bea
\label{coll}
	 \vec{v}_i^{\prime}=\vec{v}_i-\frac{1+e}{2}\left[\hat{n} \cdot\left(\vec{v}_i-\vec{v}_j\right)\right] \hat{n}, \nonumber \\
	 \vec{v}_j^{\prime}=\vec{v}_j+\frac{1+e}{2}\left[\hat{n} \cdot\left(\vec{v}_i-\vec{v}_j\right)\right] \hat{n},
\eea
where $e(<1)$ is the coefficient of restitution. Here, $\hat{n}$ is a unit vector pointing from the position of particle $j$ to the position of particle $i$.

Much like the molecular gas, we  can associate a temperature with the granular gas. This temperature, called the granular temperature, is defined as $T=\left\langle\vec{v}^2\right\rangle / d$, where $\left\langle\vec{v}^2\right\rangle$ is the mean-squared velocity, and $d$ is the dimensionality. In the early stages and in the absence of any external drive, the time rate of change of granular temperature is given by \cite{haff83}
\begin{equation}
\label{cool}
\frac{dT}{dt}=-\frac{\epsilon\omega(T)T}{d}, \quad \epsilon=1-e^2 ,
\end{equation}
where $\omega(T)$ represents the frequency of collision at temperature $T$. From kinetic theory of gases, we know that $\omega(T)$ is given by  \cite{cc70}:
\begin{equation}
\label{omega}
\omega(T)\simeq \frac{2\pi^ {(d-1)/2} }{\Gamma (d/2)}~\chi(n)nT^{1/2} ,
\end{equation}
where $\chi(n)$ represents pair correlation function at contact for hard spheres with density $n$. Using equations~(\ref{cool}) and (\ref{omega}), we arrive at the Haff's law for the HCS:
\begin{equation}
T(t)=T_0 \left[ 1 + \frac{\epsilon\omega(T_0)}{2d}~t\right]^{-2} ,
\end{equation}
where $T_0$ is the initial temperature. 
If we define the average number of collsions in time $t$ as $\tau $, then  
\bea
\tau(t) &=& \int^{t}_{0}dt'\omega(t') \nonumber \\
&=& \frac{2d}{\epsilon} \ln \left[1 + \frac{\epsilon\omega(T_0)}{2d}t\right] .
\eea
As the system looses energy, the number of collisions increases logarithmically (instead of a linear increase) with time. The Haff's law as a function of $\tau$, can be written in the following form:  
\begin{equation}
\label{haff}
T(\tau)=T_0 \exp \left(-\frac{\epsilon}{d}~\tau \right).
\end{equation}

In presence of external driving, the injected energy compensates for the loss due to collisions, the system settles to a non-equilibrium steady state.  For a driven granular system,  the stochastic equation of motion is described as,
\begin{equation}
    \dfrac{d \textbf{v}_i}{dt} = \dfrac{d \textbf{F}_i^c}{m} + \dfrac{d \textbf{F}_i^t}{m} 
\end{equation}
where $m$is the mass of the particle, $\textbf{F}_i^c$ is the force on the $i^{th}$ particle $(i = 1, 2, ...N)$ due to pairwise collision given by relation \ref{coll} and $\textbf{F}_i^t$ is the external force which is considered as Gaussian white noise with zero mean and is uncorrelated for different particles i.e,
\begin{equation}
    \langle F_{i,\alpha}^t (t) F_{j,\beta}^t (t)\rangle = \xi_0^2 \delta_{ij} \delta_{\alpha \beta} \delta(t-t') 
\end{equation}
\begin{equation}
    \langle \textbf{F}_i^t(t) \rangle = 0
\end{equation}
where $\alpha, \beta = [x,y,z]$, $\xi_0$ characterizes the strength of stochastic force, $\delta_{ij}$ and $\delta_{\alpha \beta}$ are kronecket delta and $\delta(t-t')$ is the delta function.
	
\section{\label{vdf} VELOCITY DISTRIBUTION FUNCTION}
The standard approach to study velocity distributions in the HCS is the inelastic version of the Boltzmann transport equation \cite{nb_ktgg}. In the steady state any arbitrary distribution evolves into the MB distribution:
\begin{equation}
\label{fig:MB_VDF}
P_{\mathrm{MB}}(\vec{v})=\left(\frac{1}{\pi v_0^2}\right)^{d / 2} \exp \left(-\frac{\vec{v}^2}{v_0^2}\right), \quad v_0^2=\frac{2\left\langle\vec{v}^2\right\rangle}{d}
\end{equation}
The distribution function is time-dependent due to the cooling process in the near-elastic case 
$(e \simeq 1)$. It satisfies a scaling form that deviates from the MB form\cite{gs95, vne98}:
\begin{equation}
P(\vec{v}, t)=\frac{1}{v_0^d(t)} F\left[\frac{\vec{v}}{v_0(t)}\right] \equiv \frac{1}{v_0^d(t)} F(\vec{c}).
\end{equation}
Here, $v_0^2(t)=2\left\langle\vec{v}^2\right\rangle / d$, and
\begin{equation}
\label{sonine_pe}
F(\vec{c})=\frac{1}{\pi^{d / 2}} \exp \left(-c^2\right) \sum_{n=0}^{\infty} a_n S_n\left(c^2\right).
\end{equation}
The scaled velocity distribution $F(\vec{c})$ has been expanded in terms of the Sonine polynomials in the above equation (\ref{sonine_pe}). While the Sonine polynomial expansion is a powerful tool for characterizing deviations from 
Maxwell-Boltzmann statistics, its specific form and convergence properties depend strongly 
on the driving mechanism. The coefficients presented here are specific to systems driven by 
white-noise thermostats, and different heating mechanisms may require modified expansions or 
exhibit different convergence behaviors.
For reference, a few Sonine polynomials are given by,  
\bea
S_0\left(c^2\right)&=&1, \quad S_1\left(c^2\right)=\frac{d}{2}-c^2, \nonumber \\
S_2\left(c^2\right)&=&\frac{d(d+2)}{8}-\frac{(d+2)}{2} c^2+\frac{c^4}{2}, \nonumber \\
S_3\left(c^2\right)&=&\frac{d(d+2)(d+4)}{48}-\frac{(d+2)(d+4)}{8} c^2 \nonumber \\ 
&& +\frac{(d+4)}{4} c^4-\frac{c^6}{6}, \nonumber \\
S_4\left(c^2\right)&=& \frac{d(d+2)(d+4)(d+6)}{384} -\frac{(d+2)(d+4)(d+6)}{48} c^2 \nonumber \\
&& +\frac{(d+4)(d+6)}{16} c^4 -\frac{(d+6)}{12} c^6+\frac{c^8}{24}, \nonumber \\
S_5\left(c^2\right)&=&  \frac{d(d+2)(d+4)(d+6)(d+8)}{3840} \nonumber \\
&& -\frac{(d+2)(d+4)(d+6)(d+8)}{384} c^2 \nonumber \\
&& +\frac{(d+4)(d+6)(d+8)}{96} c^4-\frac{(d+6)(d+8)}{48} c^6  \nonumber \\
&& +\frac{(d+8)}{48} c^8-\frac{c^{10}}{120}, \text { etc. } 
\eea
The Sonine polynomials, satisfy the following orthogonality relation:
\begin{equation}
\int_0^{\infty} d c c^{d-1} \exp \left(-c^2\right) S_n\left(c^2\right) S_m\left(c^2\right)=\delta_{n m} \frac{\Gamma(n+d / 2)}{2 n !} .
\end{equation}
The departure from the Maxwell-Boltzmann velocity distribution is quantified in terms of the  coefficients $a_n$ in the Sonine polynomial expansion. In the absence of dissipation, all but the leading coefficient vanish. When dissipation is turned on, all coefficients $a_n (n\geq 2)$ assume non-zero values (coefficients $a_0=1$ and $a_1=0$ in both cases). 

Using methods of kinetic theory, Brilliantov and Poschel (BP) obtained the expressions for the first two nontrivial Sonine coefficients $\left(a_{2}\right.$ and $\left.a_{3}\right)$ in HCS. For $d=3$, they obtained the following expressions \cite{bpepl2006}:
\bea
a_{2}= & -\frac{16}{c(e)}\left(-1623+1934 e+895 e^{2}-364 e^{3}\right. \nonumber \\
& \left. +3510 e^{4}-7424 e^{5}+3312 e^{6}-480 e^{7}+240 e^{8}\right), \nonumber \\
a_{3}= & -\frac{128}{c(e)}\left(217-386 e-669 e^{2}+1548 e^{3}+154 e^{4}\right. \nonumber \\
 & \left.-1600 e^{5}+816 e^{6}-160 e^{7}+80 e^{8}\right), \nonumber \\
c(e)= & 214357-172458 e+112155 e^{2}+25716 e^{3}-4410 e^{4}  \nonumber \\
& -84480 e^{5}+34800 e^{6}-5600 e^{7}+2800 e^{8}
\eea
In order to obtain the time evolution of $a_n(t)$, we use the expansion 
\begin{equation}
    \langle c^{2k}\rangle(t) = \langle c^{2k}\rangle_{\text{MB}} \sum_{n=0}^{k}(-1)^n \frac{k!}{n!(k-n)!} a_n(t)
\end{equation}
where $\langle c^{2k}\rangle_{\text{MB}} = \frac{\Gamma(k+d/2)}{\Gamma(d/2)}$. This yields the first few $a_n$ 's as follows:
\bea
a_1(t)&=&1-\frac{\left\langle c^2\right\rangle}{\left\langle c^2\right\rangle_{\mathrm{MB}}}=0, \nonumber \\ 
a_2(t)&=&-1+\frac{\left\langle c^4\right\rangle}{\left\langle c^4\right\rangle_{\mathrm{MB}}}, \nonumber \\
a_3(t)&=&1+3 a_2-\frac{\left\langle c^6\right\rangle}{\left\langle c^6\right\rangle_{\mathrm{MB}}}, \nonumber \\
a_4(t)&=&-1-6 a_2+4 a_3+\frac{\left\langle c^8\right\rangle}{\left\langle c^8\right\rangle_{\mathrm{MB}}}, \nonumber \\
a_5(t)&=&1+10 a_2-10 a_3+5 a_4-\frac{\left\langle c^{10}\right\rangle}{\left\langle c^{10}\right\rangle_{\mathrm{MB}}}, \text { etc. }
\eea
\section{\label{details_results} SIMULATION DETAILS AND RESULTS}
\begin{figure}[H]
    \centering
    \rotatebox{270}{
        \includegraphics[scale=0.55]{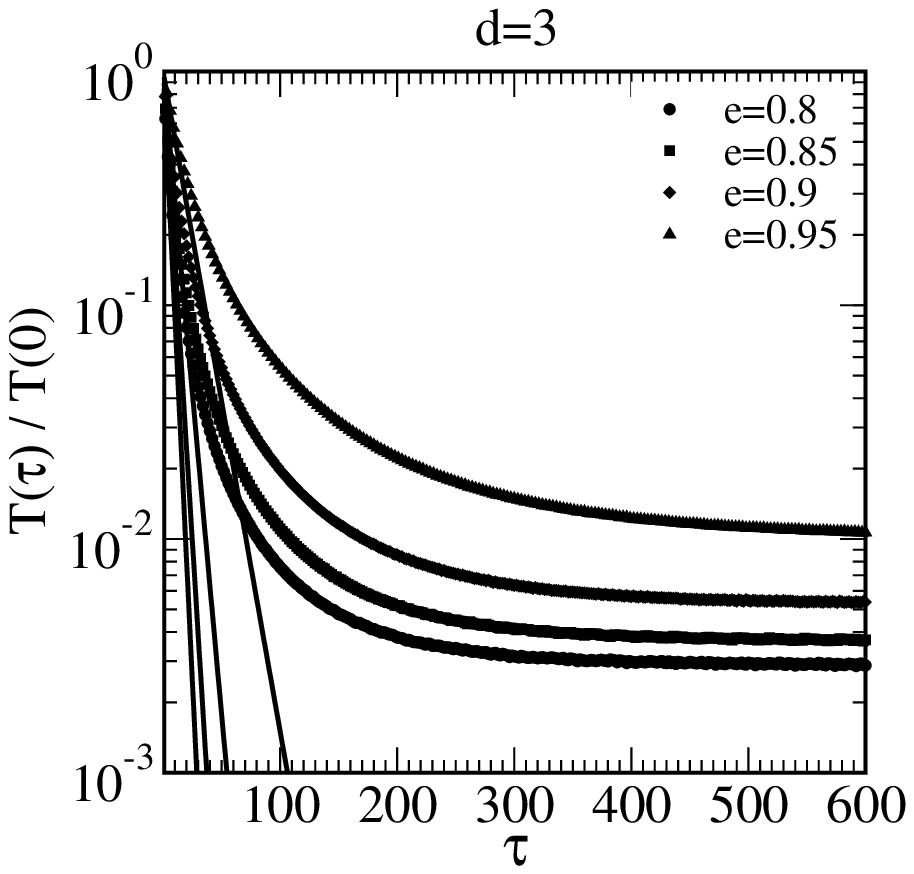} 
    }
    \caption{Time dependence of the granular temperature in $d = 3$, shown on a semilog scale. We plot the normalized granular temperature $T(\tau)/T(0)$ vs $\tau$ for $e = 0.80, 0.85, 0.90$, and $0.95$. The solid lines denote Haff's law. For the chosen noise strength $\xi_0 = 0.001$, the initial temperature $T(0)$ lies above the steady-state value, resulting in initial decay. Different initial conditions or noise strengths could lead to different early dynamics.}
    \label{fig_haff1}
\end{figure}

The system is initialized by assigning each particle a random position and velocity. Our system consists of $N=500000$ particles confined in a 3D cubical box with periodic boundary conditions such that number density is $n=0.02$. The initial position is such that the core of no two particles is allowed to overlap. The random velocity components are so chosen that $\Sigma \vec{v}_i = 0$. The system is then evolved to $\tau = 100$ at $e = 1$ without any input of energy. This ensures that the system has relaxed to a MB velocity distribution. This serves as the initial condition for our simulation. We then evolve the system till $\tau = 1000$ for four different values of $e$ ($e = 0.95, 0.90, 0.85$ and $ 0.80$). Results presented here correspond to averages over $50$ independent initial conditions.

We used event-driven MD to simulate a system of hard sphere inelastic particles \cite{allentild, rapaport}. All particles are identical with unit mass $m=1$ and diameter $\sigma=1$. The postcollision velocities are obtained from precollision velocities by the relation \ref{coll}. The system is subject to Gaussian white noise where the heat component is added to the velocity of each molecule after every time step $dt$ as follows,
\begin{equation}
    v_i(t+dt) = v_i(t) + \sqrt{r} \sqrt{dt} \xi. 
\end{equation}
Here $\xi$ is random variable which is uniformly distributed between $\left[ -1/2, 1/2 \right]$, r is amplitude of noise chosen to be 0.001. After adjusting the velocities, the system is shifted to centre of mass frame, to ensure conservation of linear momentum. 
\begin{equation}
    \textbf{v}_i = \textbf{v}_i - \dfrac{1}{N} \sum _{i=1}^{N} \textbf{v}_i. 
\end{equation}

The evolution of the system's kinetic energy is governed by two competing processes: energy dissipation through inelastic collisions (characterized by the coefficient of restitution $e$) and energy input from the white noise thermostat (characterized by the noise strength $\xi_0$). The initial dynamics depend on the relative strength of these processes and the initial temperature $T(0)$. When the initial kinetic energy exceeds the steady-state value determined by $e$ and $\xi_0$, the system exhibits temperature decay as energy loss through collisions dominates. Conversely, if the initial energy is below the steady-state value, the system's temperature increases as energy input from the thermostat exceeds dissipation. Eventually, these competing effects balance, leading to a non-equilibrium steady state.

\begin{figure}
    \centering
	\includegraphics[scale=0.55]{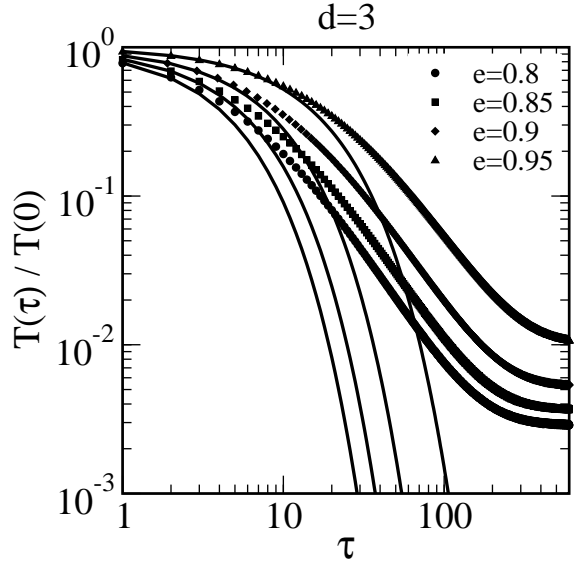} 
    \caption{Time dependence of the granular temperature in $d = 3$, shown on a log-log scale. It can be noticed that the temperatures for each value of $e$ settle into a constant value, representing the balance between energy input from the thermostat and dissipation through collisions.}
    \label{fig_haff2}
\end{figure}

In Fig. \ref{fig_haff1}, the time evolution of the reduced temperature $T(\tau)/T(0)$ as a function of $\tau$ on a semilog scale is plotted for different values of restitution coefficient $e$. For reference, Haff's law is also plotted in solid lines. For our chosen parameters ($\xi_0 = 0.001$ and initial conditions), the system shows initial decay before reaching steady state. Fig. \ref{fig_haff2} shows the time evolution of the reduced temperature $T(\tau)/T(0)$ as a function of $\tau$ on a log-log scale for $e = 0.8, 0.85, 0.9$ and $0.95$. 

\begin{figure}
    \centering
	\includegraphics[scale=0.5]{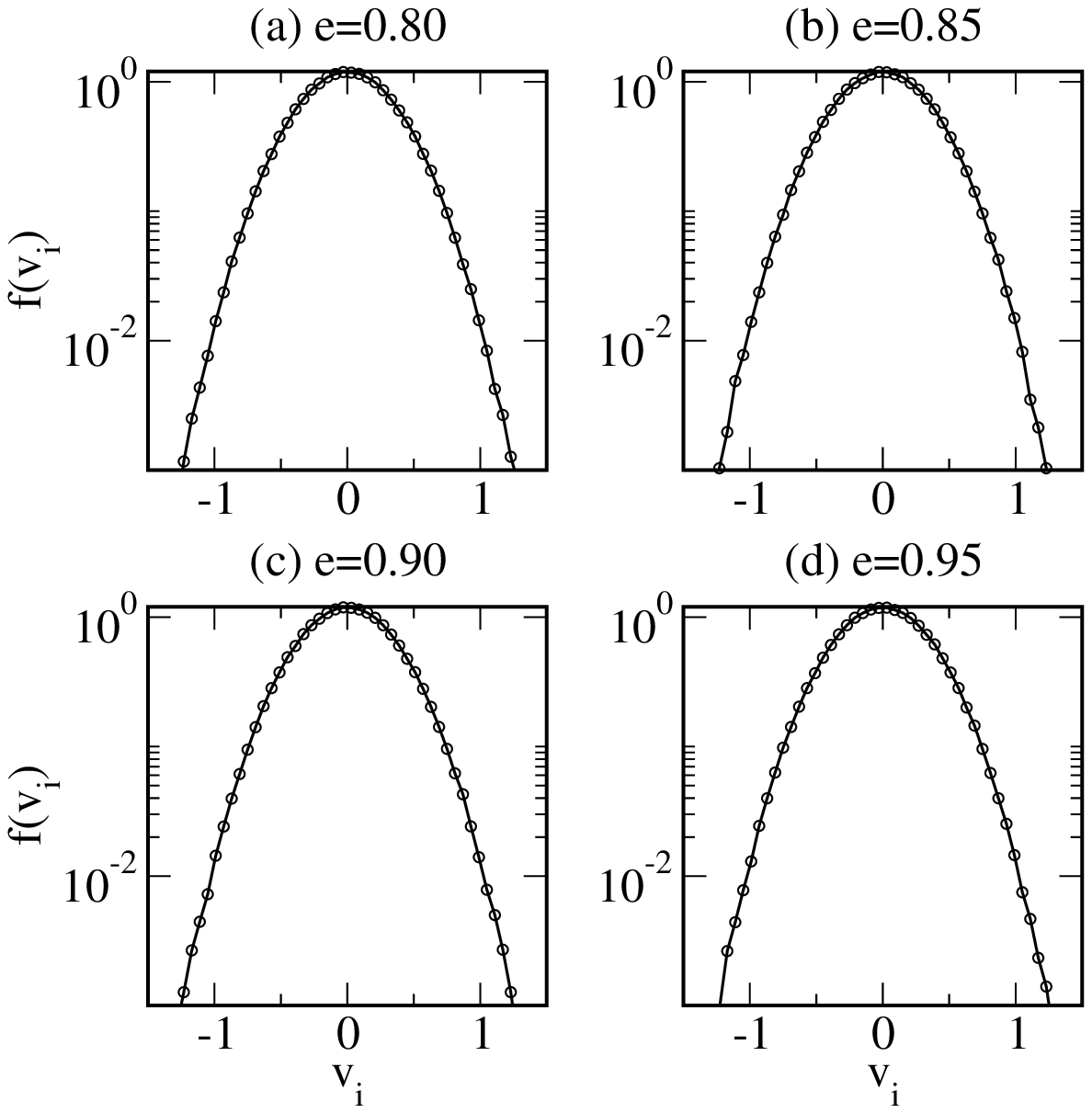} 
    \caption{Plot of the steady state velocity distribution functions $f(v_i)$ for different 
    values of $e$. Plots \textbf{(a), (b), (c)} and \textbf{(d)} correspond to $e = 0.80, 0.85,
    0.90$, and $0.95$ respectively. The solid line in each figure represents the scaled MB 
    distribution for the corresponding steady state. Circles represent results obtained from our 
    numerical simulation.}
    \label{fig_VDF}
\end{figure}

Next, we present the results for the steady state velocity distribution function (VDF). An arbitrary velocity distribution evolves into the Maxwell-Boltzmann VDF (see Eqn. (\ref{fig:MB_VDF})). Figure \ref{fig_VDF} shows the VDFs for different values $e$. Numerical details are given in the figure caption. The data obtained from our simulation shows slight deviation from the MB distribution in the steady-state. 

\begin{figure}
    \centering
	\includegraphics[scale=0.7]{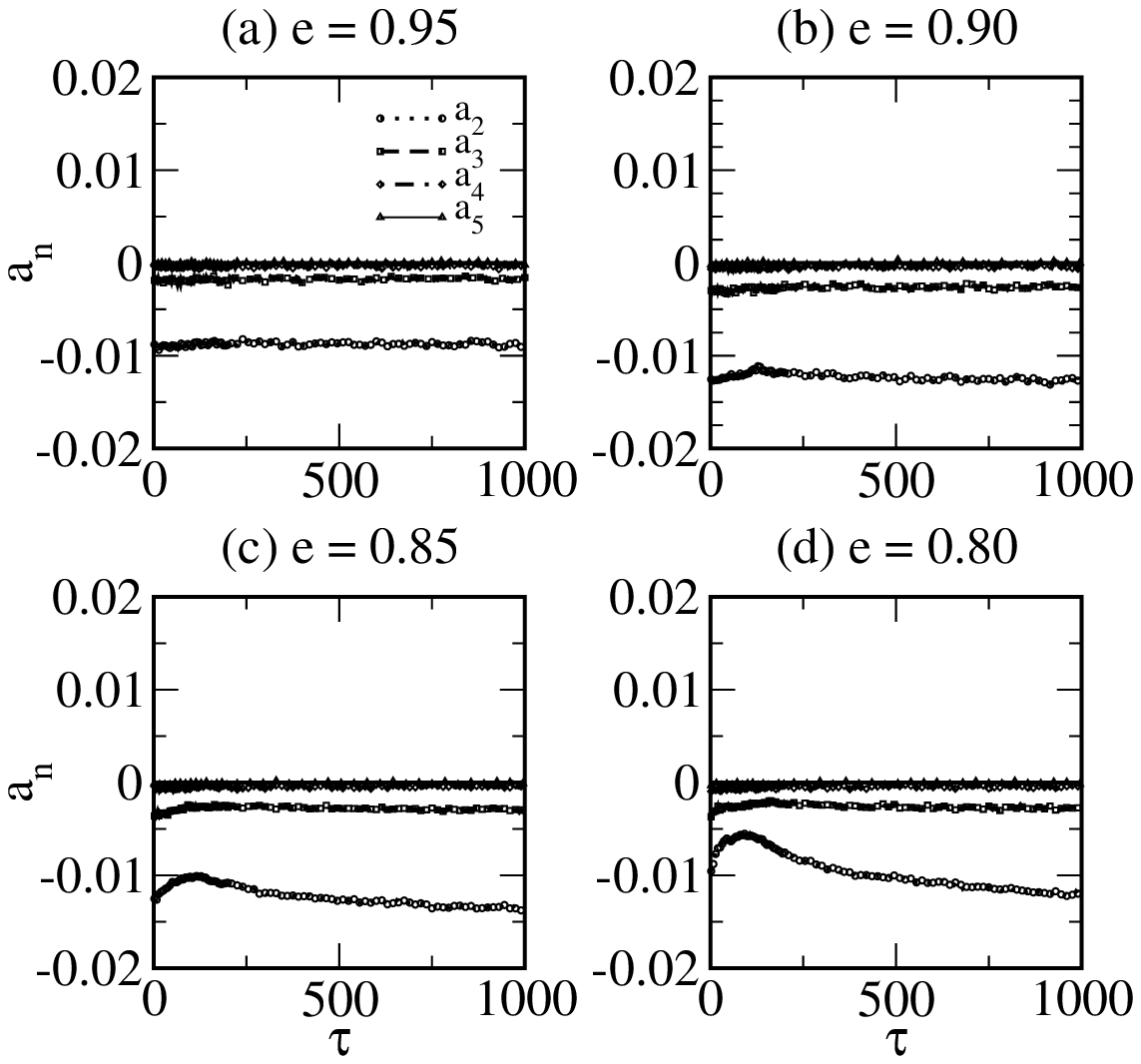} 
    \caption{Time-evolution of $a_2$, $a_3$, $a_4$ and $a_5$ for different values of $e$: (a) $e = 0.95$, (b) $e = 0.90$, (c) $e = 0.85$ and (d) $e = 0.80$. The results presented here are averages over 50 independent runs.}
    \label{fig_sonine_coeff}
\end{figure}

Next, we study the time evolution of the coefficients of the Sonine polynomial expansion. The deviation from Maxwell-Boltzmann VDF is characterized by non-vanishing values of the coefficients $a_k, k\geq2$. In fig. \ref{fig_sonine_coeff}, we plot the Sonine coefficients $a_2, a_3, a_4, a_5$ vs. $\tau$ for (a) $e = 0.95$, (b) $e = 0.90$, (c) $e = 0.85$ and (d) $e = 0.80$ respectively. We can clearly see that the Sonine coefficients for all values of $e$ settle to non-zero value. Successive order coefficients are found settle to smaller and smaller values, confirming the convergence of the series expansion. 

\section{\label{sec:summry} SUMMARY AND CONCLUSION}
We conclude this paper with a summary and discussion of our results. We have studied the 
dynamical properties of heated granular fluids using large-scale molecular dynamics simulation 
in three dimensions. In our MD simulations, we have explored the time evolution of granular 
temperature and the coefficients of Sonine polynomial expansion of the velocity distribution 
function of a uniformly heated granular gas. We use white-noise thermostat to compensate for 
the loss of energy due to dissipative interactions between particles. In the early stage of 
evolution, the system loses energy with time. The interplay between loss of energy due to 
inelastic interactions and energy input from the thermostat results in the system attaining 
a steady state temperature at later stages of evolution.
We tracked the system's evolution to steady state and analyzed the coefficients of the Sonine 
polynomial expansion of the velocity distribution function. A departure of the velocity 
distribution from the Maxwell–Boltzmann (MB) distribution is characterized by non-zero values of Sonine coefficients $a_n$, ($n \geq 2$). In our simulations, the Sonine coefficients 
$a_2 - a_5$ have been calculated numerically and have been found to settle to non-zero values. 
We also noticed that the successive-order Sonine coefficients are much smaller in magnitude. 
The decreasing magnitude of higher order Sonine coefficients suggests the convergence of 
Sonine polynomial expansion.
This study demonstrates the behavior of uniformly heated granular gases specifically under 
white-noise thermostating conditions. Several key limitations and considerations should be noted. The reported Sonine coefficient values are specific to white-noise thermostats and may not generalize to other driving mechanisms.
 Different thermostating methods (e.g., boundary driving, velocity scaling, or deterministic 
thermostats) could lead to:
 Different steady-state velocity distributions,
Modified forms of the Sonine expansion,
Different convergence properties of the expansion coefficients.
The convergence of the Sonine expansion observed in our system is not guaranteed for 
other driving mechanisms. Our results provide a baseline for understanding velocity statistics in white-noise thermostated granular gases.
Future work could explore:
Comparative analysis with other thermostating mechanisms,
Investigation of system-size dependence of the Sonine coefficients,
Extension to more realistic driving mechanisms that might better represent experimental conditions,
Development of theoretical frameworks that can predict thermostat-dependent variations in 
velocity statistics.

\section*{\label{ack} ACKNOWLEDGEMENTS}
RFS acknowledges financial support from University Grants Commission in the form of Non-NET fellowships. He also wishes to acknowledge the computational facilities at the Department of Physics, JMI.

\end{document}